# Reflecting on Creative-Boundaries with an AI Co-Doodler

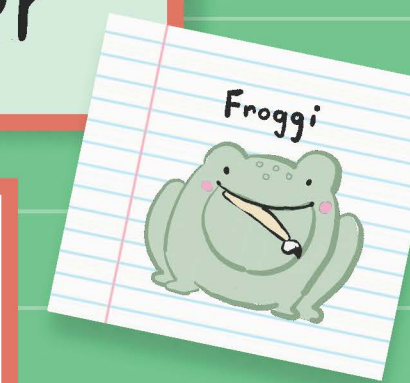

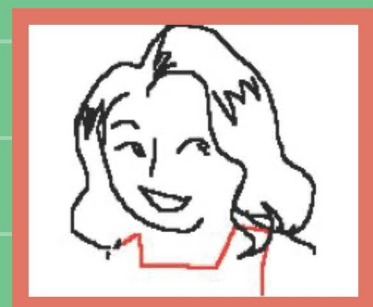
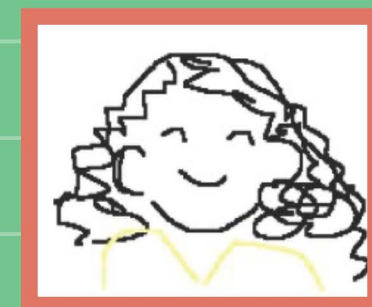
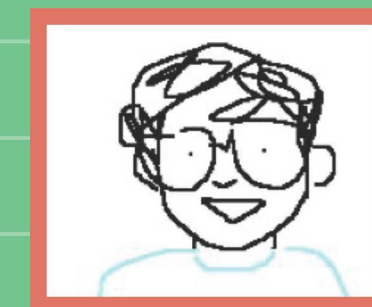
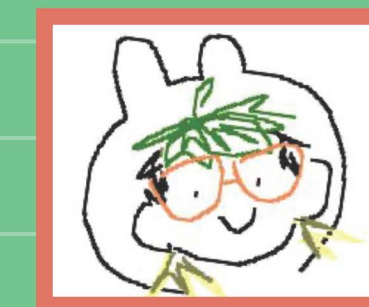

**Samia Menon**
UC Berkeley
Berkeley, California, USA
**samia_menon@berkeley.edu**

**Samyukta Jayaram**
UC Berkeley
Berkeley, California, USA
**samyukta.jayaram@berkeley.edu**

**Chetan Goenka**
UC Berkeley
Berkeley, California, USA
**cgoenka@berkeley.edu**

**Shm Garanganao Almeda**
UC Berkeley
Berkeley, California, USA
**shm.almeda@berkeley.edu**

## ABSTRACT

In this pictorial, we consider how the negotiation of creative boundaries with a co-creative AI system can create moments for personal creative reflection. We ground this in our experiences with Froggi-Draw, a single-initiative co-doodling system that gives users power to decide when and how much an AI "collaborator" (Froggi) contributes to their drawing. From a 1-week pilot study where (n=8) novice and experienced artists doodled daily with the system, we report on ways participants navigated creative risk and uncertainty, and how their usage and perceptions of Froggi shifted over time. In moments of disruption, participants described the system as encroaching on their *creative territory*. We consider how the design of a supportive, co-creative AI "collaborator" might look like the design of a supportive *power dynamic*—and finding ways to offer users control to find, reflect upon, and flexibly negotiate the boundaries of that dynamic.

**Reference:**
Samia Menon, Samyukta Jayaram, Chetan Goenka, and Shm Garanganao Almeda. 2026. Reflecting on Creative Boundaries with an AI Co-Doodler. In *Proceedings of The First Reflection in Creative Experience (RiCE) Workshop (RiCE W1).* ACM Creativity & Cognition 2026, London, UK.

## INTRODUCTION

Many creative domains are seeing the introduction of co-creative AI "collaborator" systems [1]. Given that even well-intentioned co-creative system interventions can be disruptive, or even harmful[2, 3], HCI researchers are increasingly interested in understanding how to design AI systems that meaningfully support creative processes.

We are interested in how the negotiation of creative **boundaries** with a co-creative system can necessitate, and create moments for, personal creative reflection. We ground this position in our experiences developing and studying **Froggi-Draw:** a single-initiative, co-doodling system that provides users with the power to decide when, and how much, they would like the system to intervene in their drawing.

We conducted a pilot user study with (n=8) experienced artists and novices who made doodles with the system daily for 1-week. Through analyzing their daily reflections and responses to semi-structured exit interview questions, we identified themes in how their use and perceptions of the system shifted over time. Participants learned to thoughtfully negotiate the **risk** and **uncertainty** involved when inviting Froggi to intervene in their drawings. In moments of disruption, participants described the system as encroaching on their creative territory: feeling the system had taken up too much (literal or conceptual) space. Our findings led us to consider how the task of designing a supportive, co-creative 'collaborator' may resemble the process of designing a supportive power dynamic between user and system, and **offering the user control to flexibly reflect upon, and negotiate, the boundaries of that dynamic.**

In this pictorial, we present the Froggi-Draw system and findings from our pilot study, along with relevant themes surfaced by our preliminary data analysis.

# RELATED WORK

Zhou & Sterman [10] found that generating **intermediate** text encouraged rewriting facilitating more engagement from human users in a co-creative writing process. Intermediate forms of drawing like sketching and doodling are significant mediums for visual communication [4, 7, 8, 9] and for autotelic [5, 6] creative expression, making it a well-motivated domain for exploring co-creative interaction design. However, few studies of co-creative drawing systems have focused on investigating the impact these systems have on the creative experience. Notably, Lawton et. al develop **Reframer** [11], which supports context-aware, mixed-initiative, synchronous drawing with an AI-system. Their "Model of Co-Creative Agentive Flow" framework describes how users moved between treating the system as a collaborator and a tool—"as they adapt to and are disrupted by a tool's functions, with different features helping at some points in the process and hindering at others." Lawton et al. also found that the mixed-initiative co-creativity experience can be "unfamiliar" and "confronting.

Rather than tackling research questions of mixed-initiative creativity support design (**When would a good collaborator intervene into the creative process?**), we were interested in what we could learn from a **single-initiative interaction** where both user and system operated at the same **intermediate drawing** quality. Inspired by human-human co-creative drawing games like "Pass the Canvas", we designed a drawing interaction where users had the power to **decide** when to ask for AI intervention while collaborating on a doodle —allowing us to ask...

- **When and how might we relinquish control to an AI "collaborator"?**
- **What can interactions with an AI "collaborator" teach us about how human creatives perceive and relate to these systems?**
- **What can interactions with an AI "collaborator" teach creatives about themselves?**

P4: "When Froggi contributed more, I thought it was cool but even when I add to it, it doesn't really feel like mine...."

Froggi-Draw was also designed to be a more "reflective" system than other co-doodling tools. While many existing tools rely on next-stroke prediction, our system constructs an entire image "vision" from which it samples strokes!

See future work for our implementation!

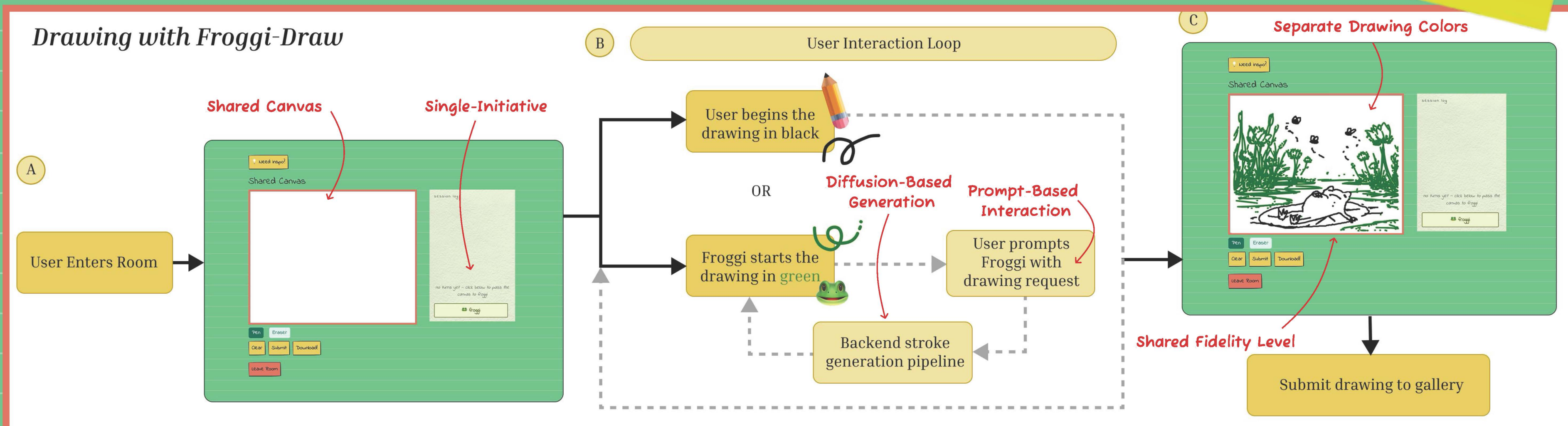

# Froggi-Draw Encouraged User Reflection

While experimenting with Froggi-Draw, users quickly found that the unique characteristics of the system urged them **to reflect on their creative boundaries**. In other words, users had to understand what aspects of a doodle **made it their own** and what aspects they could **relinquish** to the system. We uncovered themes by analyzing users' responses to **open-ended daily prompts** ("Share any notes on what you drew today", "Describe your relationship with Froggi") and final **semi-structured exit interviews**.

There were a few characteristics that particularly stood our for their role in urging users to reflect on their ***creative boundaries***

Need inspo?

## Shared Canvas

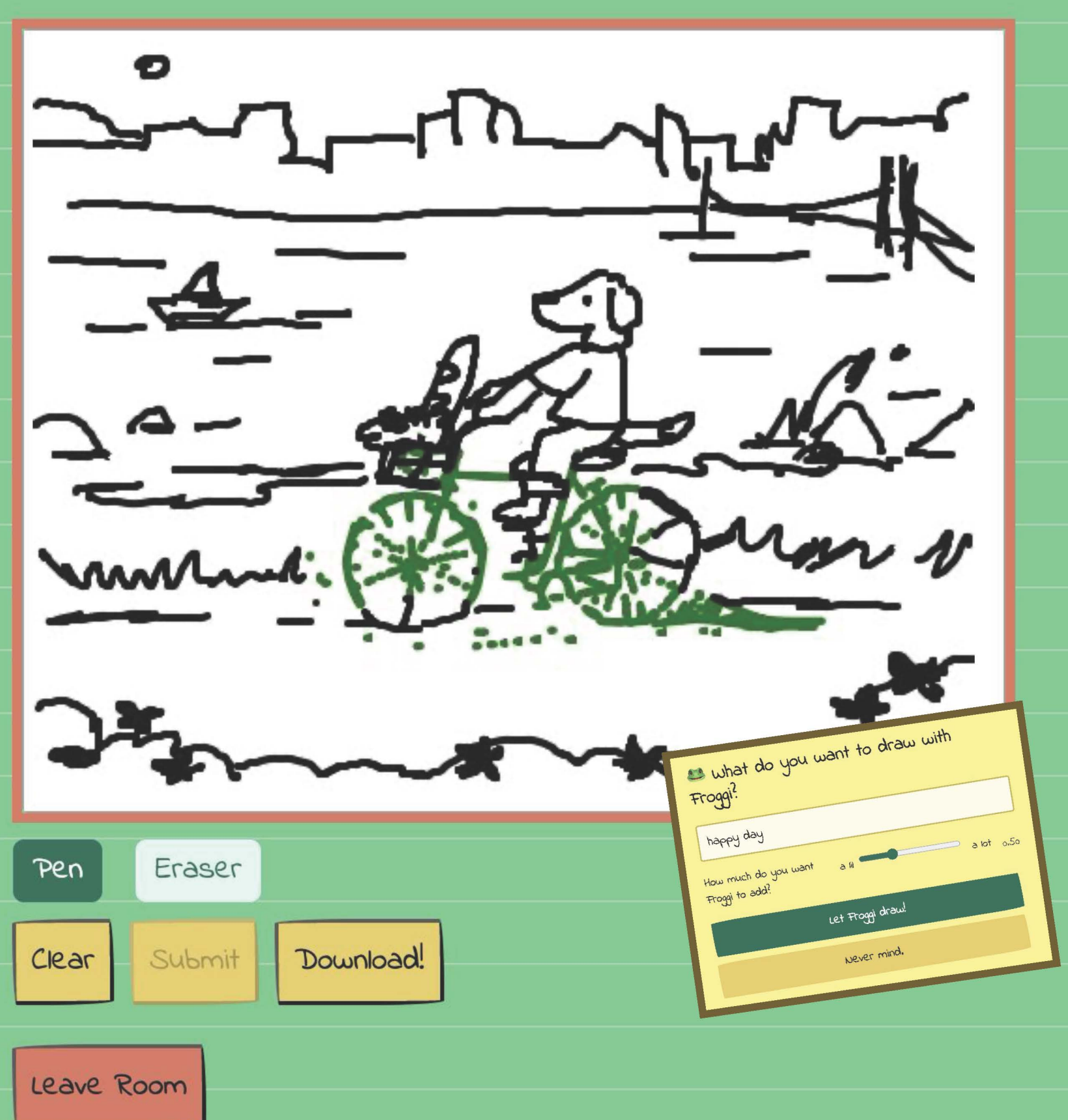

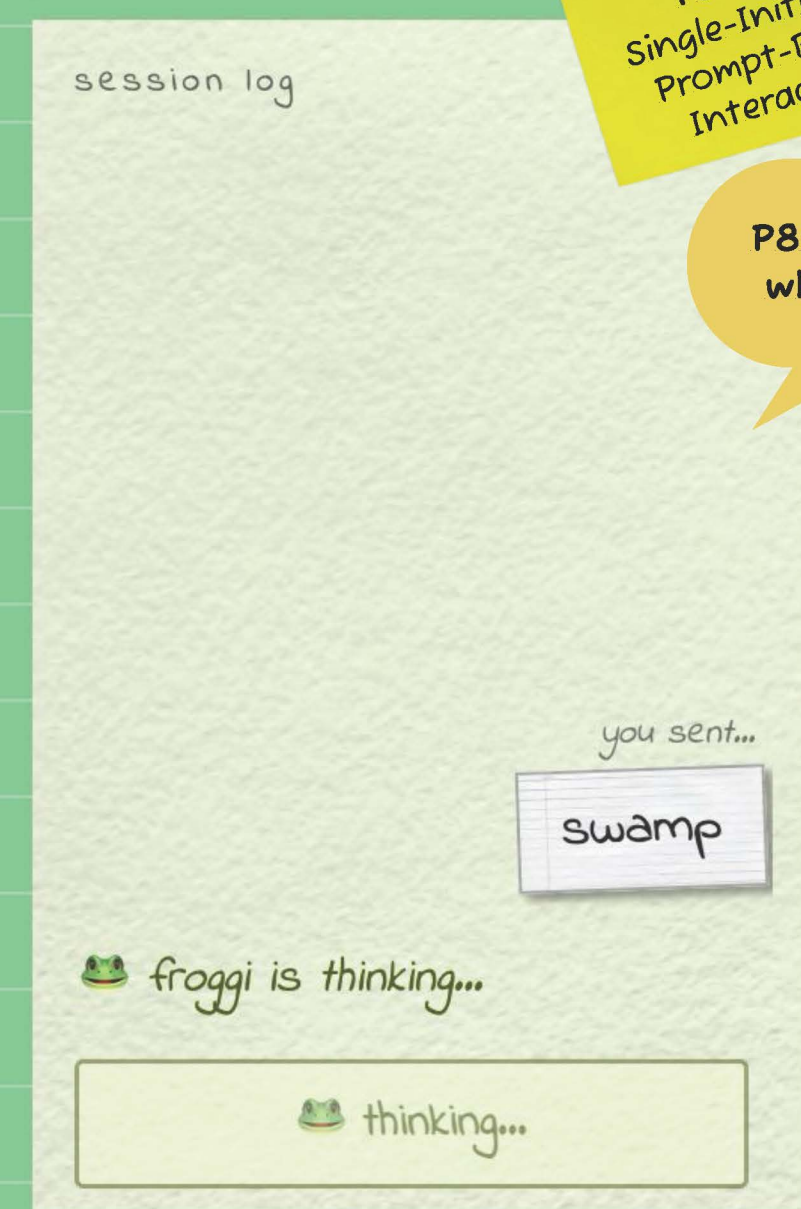

## Froggi **undermines human-human collaborative norms** which encouraged reflection on **what boundaries were crossed**

**P4: "When froggi is just adding something in the middle, and I can't do anything about it, it felt frustrating and disrespectful sometimes"**

**P3: "[My relationship with Froggi is] Combative"**

**Enabling Design Features:** Shared Canvas, Shared Fidelity Level

## Froggi facilitated thoughtful **articulation** which encouraged reflection on **their ultimate creative goal**

**Enabling Design Feature:** Single-Initiative, Prompt-Based Interaction

**P4: "I think prompting more ambiguously was more important for me than for the system"**

**P8: "I think I need to be more specific when telling the model what to add"**

## Froggi generated **unpredictable outputs** which encouraged reflection on **personal flexibility**

**P1: "I had a clear idea but changed because of what froggi added, but that felt like a happy accident."**

**P4: "The fact that Froggi's unpredictability impacts the composition and contents of a piece make it feel like a bigger or more meaningful course change in the work."**

**Enabling Design Feature:** Unpredictability of Diffusion Models

**P6: "I think in creative tasks it is useful, like arts, or tasks where you don't necesarrily[sic] have a high risk workflow. It just makes stuff more interesting....I would feel the most comfortable using a tool like this for exploration"**

https://froggi-draw.vercel.app/

Once users were able to reflect on their creative boundaries, they developed interaction strategies to maintain them!

## Creative Challenger

We characterize some use-strategies as positioning Froggi-Draw as a **creative challenger**: Steering the system towards more indefinite and suggestive outputs, with the goal that its generations leave ample space for the human user to shape the final output and lead its creative direction. P4 used a low sampling rate to steer the system to contribute a small number of abstract strokes, allowing them to shape the scene around it. In this way, users used the system to prompt their own creativity and **maintained a stronger sense of ownership**.

## Tool

The most straightforward approach to using Froggi-Draw is using it as a **tool**: the user has a clear intention of what they want the system to add to the drawing. P5 explicitly requested Froggi to draw leaves growing from a boot (see Fig. bottom-right) to complete the drawing. For our study participants, choosing an "optimal" interaction strategy involved balancing leveraging the expressivity inherent to a generative, open-ended system, with a sense of maintaining creative ownership and control. **Users expressed constraints that they wanted Froggi to adhere to.**

## Scaffold/Teacher

Several participants used Froggi-Draw's outputs as **creative scaffolding**. P8 prompted the system to draw a bird (see Fig bottom-left), then mimicked the style of the generated output with their own additions to the canvas. Since the system frequently rendered only part of a scene, participants were inclined to complete the sketch in a similar visual language. P3 adopted a pointillist quality to match the style of a partially-generated sketch. **Users articulated where they wanted support in achieving their creative goals with system guidance.**

### Creative Challenger

P4:

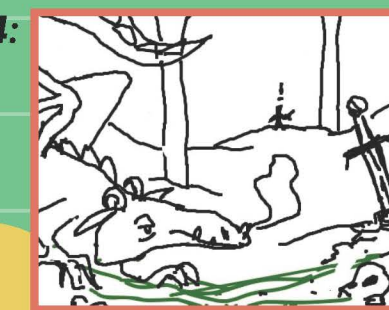

**P4:** "I decided to just use what Froggi outputted more as inspiration"

**P3:** "Froggi has been more helpful as a jumping off point than a contributor to an already developed drawing for me."

P2:

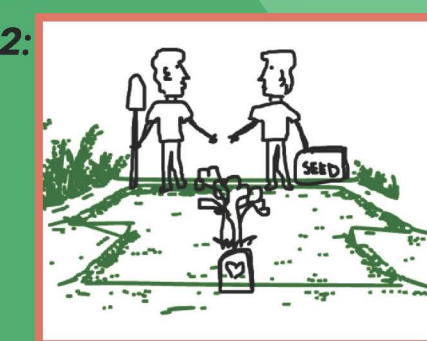

**P1:** "I had a clear idea but changed because of what froggi added, but that felt like a happy accident"

P5:

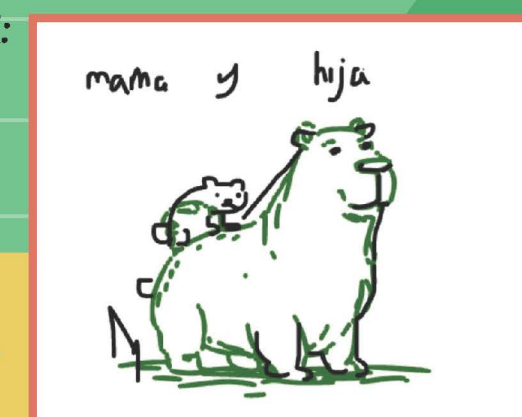

**P8:** "Froggi is my teacher"

P1:

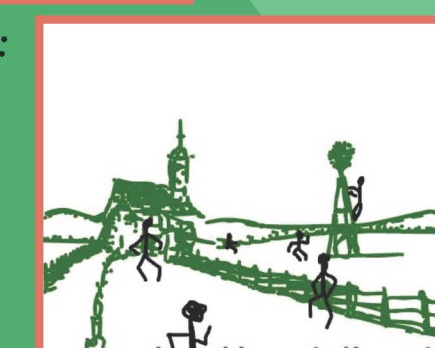

**P3:** "I decided to allow froggi to draw the part of the doodle that I knew I would struggle with accurately depicting. I think it worked"

P8:

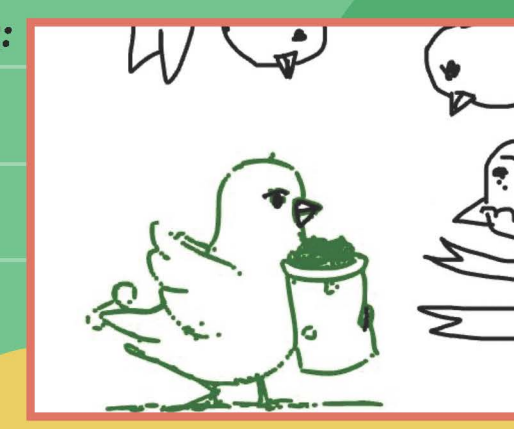

P6:

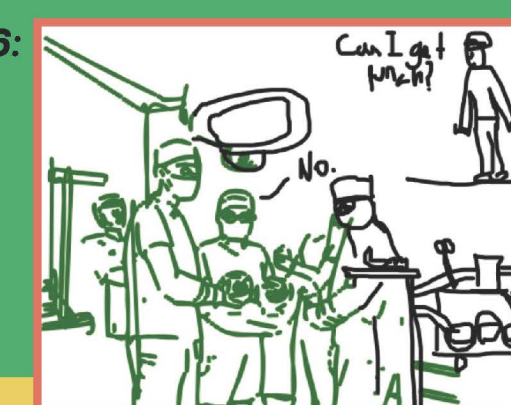

P5:

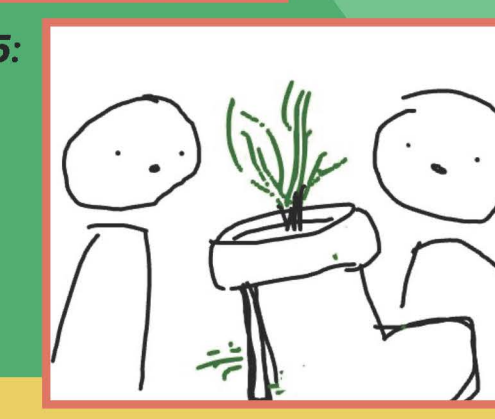

**P5:** "This time I used froggi first to develop the image and then drew based off of what I was provided. It drew half of it for me and then I filled in the rest"

**P4:** "I figured a common use case would be to get help drawing parts that aren't fun to me (like a bicycle) but might be part of the artist's vision"

### Scaffold

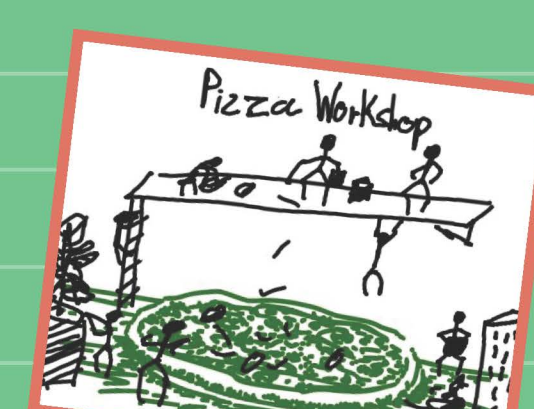

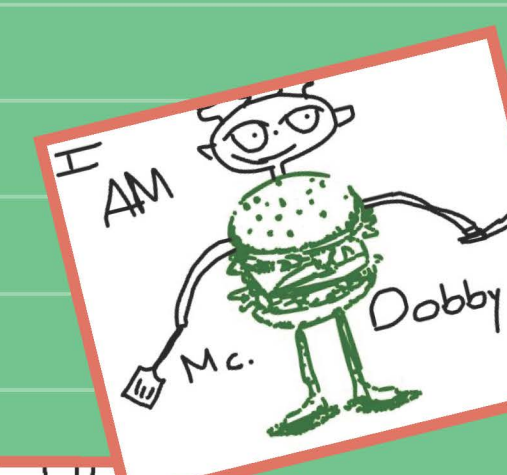

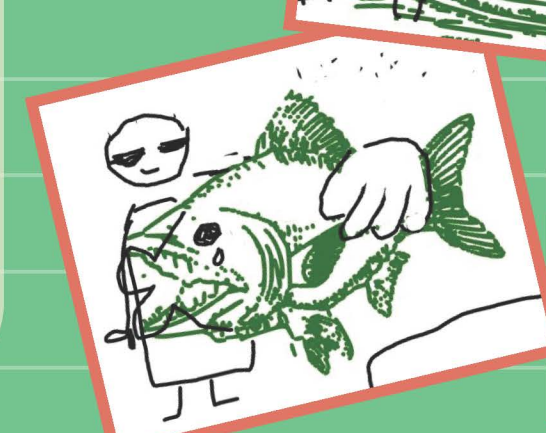

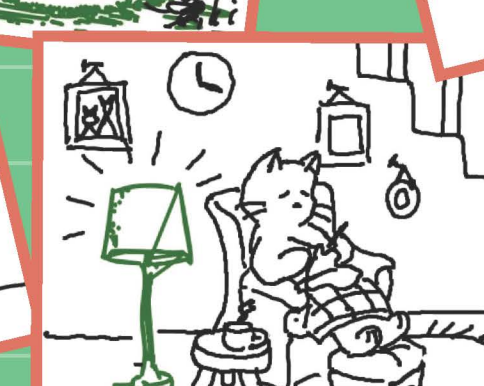

### Tool

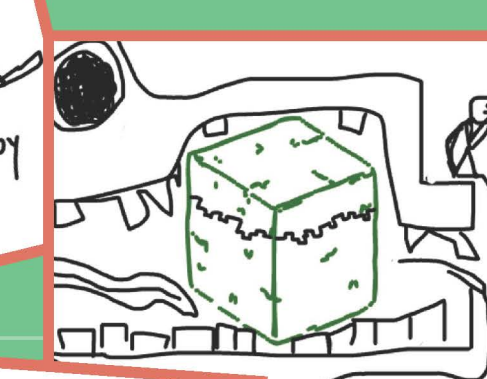

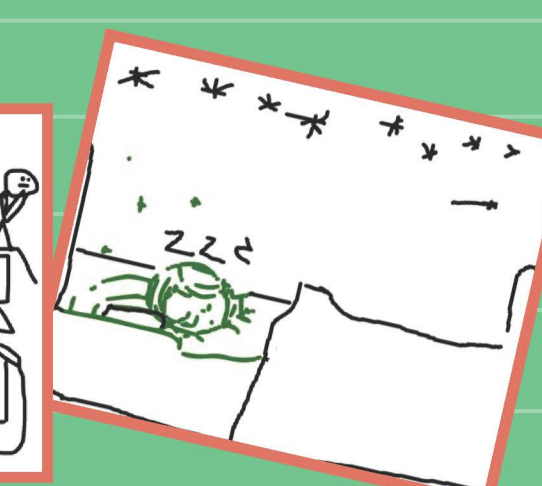

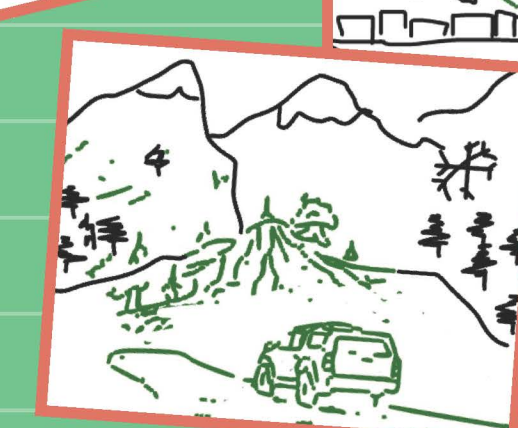

Thank you!
(and keep doodling!)

Leave Room